# Hemodynamic Performance and Blood Damage of the Intra-Aortic Pumps: A CFD-Based Investigation

**Osman Aycan[1,2], Yeojin Park[1], Lyes Kadem[1]**

[1]Laboratory of Cardiovascular Fluid Dynamics, Department of Mechanical Industrial and Aerospace Engineering, Concordia University, Montreal, QC, Canada

[2]Department of Mechanical Engineering, Faculty of Engineering, Zonguldak Bulent Ecevit University, Zonguldak, Türkiye

**Correspondence**

Osman Aycan, Department of Mechanical Engineering, Faculty of Engineering, Zonguldak Bulent Ecevit University, Zonguldak, Türkiye.

Email: osman.aycan@beun.edu.tr; lcfd@encs.concordia.ca. Tel: +90 (372) 291 1144

## Abstract

Three intra-aortic blood pumps were evaluated and compared using CFD simulations. The aim of this study is to evaluate the hemodynamic performance and calculate the hemolytic potential of the pumps. The flow fields generated by the pumps were simulated using CFD with a wall-modeled large eddy simulation (WMLES) approach. The simulations produced pressure-flow curves, hydraulic efficiency, shear stress distributions, and hemolysis predictions. A grid study was conducted using the Grid Convergence Index (GCI) method, and a new dimensionless parameter, the *Hemolytic Number (HN)*, was introduced as a standardized metric to compare hemolysis and universal pump performance. The impeller-driven pump had the highest-pressure head (~800 Pa at 4 L/min) and the best hydraulic efficiency (~6% at 14 L/min), outperforming both the single (maximum 2.7%) and triplet (maximum 2.2%) pumps. The NIH values were also lowest for the impeller pump (NIH ≈ 0.0035 g/100L), indicating high hemocompatibility. Both the single and triplet pumps showed regions of recirculation, particularly at lower flow rates. A smaller HN indicates better hemocompatibility; for the impeller-driven pump, HN remains below 1 across the investigated flow rates. Overall, the impeller-driven pump outperformed the other designs in terms of both hemodynamic and hemolytic measurements. The findings provide valuable insights for the future development of intra-aortic pumps and the personalized selection of devices for individual patients.

## 1. Introduction

Cardiovascular diseases remain the leading cause of mortality globally, and heart failure is responsible for a significant proportion of these fatalities. The growing incidence of heart failure has prompted the evolution of mechanical circulatory support (MCS) devices, specifically intra-aortic blood pumps, which offer less invasive approaches compared to conventional ventricular assist devices (VADs) [1]. Intra-aortic pumps are intended for treatment in patients suffering from acute decompensated heart failure with concomitant cardiorenal syndrome. They are intended mainly to lower cardiac afterload and increase kidney perfusion. Devices, such as Aortix™ (*Procyrion, Inc., Houston, Texas, USA*), ModulHeart™ (*Puzzle Medical Device Inc., Montreal, Canada*), and Second Heart Assist devices (*Second Heart Assist Inc., Salt Lake City, UT, USA*), are technological advances used to enhance cardiac output and facilitate end-organ perfusion in patients with diminished cardiac function and severe heart failure, thereby providing hemodynamic support and improving quality of life [2–4]. Their design and optimization require a comprehensive understanding of their hydraulic performance and hemocompatibility, with special regard to hemolysis.

Computational Fluid Dynamics (CFD) has emerged as an indispensable tool in the development and evaluation of blood pumps, enabling thorough investigations of flow fields, shear stress distributions, and the potential for blood damage associated with such devices [5]. Recent work has shown that CFD can be used not only to design novel systems, such as a pediatric pump–lung ECMO device optimized for low pressure drop, reduced hemolysis, and efficient gas transfer [6], but also to compare the flow characteristics and hemolytic performance of clinical centrifugal pumps under ECMO-relevant conditions [7, 8]. These studies typically combine CFD-predicted shear and hemolysis indices with in vitro blood-loop testing, highlighting the value of integrated computational–experimental approaches for assessing pump biocompatibility. In addition to conventional finite-volume solvers for the Navier–Stokes equations, advanced numerical frameworks such as lattice Boltzmann methods (LBM) and immersed boundary methods (IBM) have been proposed for complex fluid–structure interaction and flows with moving or flexible boundaries. Recent flexible forcing immersed boundary–LBM schemes have, for example, been applied to two- and three-dimensional fluid–solid interaction problems [9, 10] (Dash, 2019; Sikdar et al., 2023). In the present study, we focus on a finite-volume solver with the Wall-Modelled Large Eddy Simulation (WMLES) framework, while recognizing that LBM/IBM-based approaches represent a promising direction for future work. Earlier investigations have noted the significant role of impeller design, blade geometry and configuration, and rotational speed-dependent flow rates on hemodynamic efficiency and hemolysis rates in blood pumps [11]. The integration of CFD in the design process has already been instrumental in the optimization of axial and centrifugal pump geometries to minimize hemolysis [12–18]. Wu et al. (2024) provides an example of turbulence-resolving simulations in centrifugal blood pumps and discusses the importance of grid quality and near-wall treatment for capturing secondary flows and performance trends by applying LES [19].

Intra-aortic assist pump design space and clinical context. Intra-aortic pump devices are typically deployed in the descending thoracic aorta (the proximal abdominal aorta above the level of the renal arteries at the level of the T11 vertebral body) [20, 21] and provide circulatory support without crossing the aortic valve. Existing concepts span multiple design archetypes. One approach is catheter-deployed axial/entrainment intra-aortic pumps, which accelerate blood within the aorta to augment distal perfusion and reduce ventricular loading; these systems have been investigated in first-in-human studies for temporary support in HR-PCI (high-risk percutaneous coronary intervention) and in ADHF (acute decompensated heart failure) settings where cardiorenal hemodynamics are clinically relevant [2, 22]. In parallel, multi-pump configurations [21, 23] have been proposed to distribute loading across multiple rotors (e.g., multi-micro-pump concepts evaluated for cardiorenal support in HR-PCI), highlighting that both serial and parallel multi-unit strategies are being explored to balance flow augmentation and hemocompatibility [24–26]. Another approach includes propeller/impeller-based intra-aortic pumps positioned in the descending aorta, which generate a pump-induced pressure gradient and have also been clinically evaluated for HR-PCI support [27, 28] (Kapur et al., 2021; Okamoto and Mitamura, 2024).

In the context of intra-aortic blood pumps, devices such as Aortix and ModulHeart have undergone tests to determine their effectiveness and safety. Cowger et al. [2] reported the Aortix device for acute decompensated heart failure cases, and Georges et al. [23] experimented with organ blood flow against the ModulHeart support device, generating pertinent clinical data to enhance computational findings. The device position and flow dynamics are important parameters in regulating both hemodynamics and intra-aortic pump-related hemolysis. Li et al.'s [29] study examined the effects of an intravascular pump positioned in the aorta, accounting for the hemodynamic environment for various positions and numbers of pumps (n = 1 to 3). A study by Park et al. [30] used CFD to analyze the flow induced by a transcatheter intra-aortic entrainment pump and concluded that high rotational velocities can lead to severe blood damage and elevated arterial wall stress.

Since intra-aortic pump designs are varied, comparative analyses are essential to identify their strengths and limitations. The Aortix device, *a single pump design*, demonstrated effectiveness in augmenting renal perfusion and cardiac output [2]. The ModulHeart system, *a triplet pump design*, allows modular support based on patient needs [3, 23]. The Second Heart Assist device, on the other hand, has *an impeller-driven design* for providing blood flow support [28].

Despite shared goals, these designs differ in their flow-generation mechanism and constraints. Because intra-aortic devices operate in a non-occlusive configuration and interact with a confined aortic lumen, they often face trade-offs among achievable flow augmentation, local wake/recirculation losses, anchoring/positioning requirements, and hemocompatibility (shear-induced blood damage). Additionally, published studies are commonly device-specific and use different modeling assumptions, which complicates direct comparisons across architectures and operating points.

The present study uses the following three types of intra-aortic pump geometry: single-pump, triplet-pump and impeller-driven pump. These represent the Aortix™, ModulHeart™ and Second Heart Assist devices, respectively. To the best of the authors' knowledge, there is no systematic computational comparison of different intra-aortic blood pump configurations operating in the descending aorta and evaluated under a common numerical framework. In particular, the hydraulic efficiency, recirculation behavior, and hemocompatibility of single, triplet, and impeller-driven intra-aortic pumps have not been quantified side-by-side using high-fidelity simulations. Moreover, existing hemolysis metrics do not provide a dimensionless, device-independent measure that links flow generation, inertial loading, and blood damage. The present work addresses these gaps by: (i) performing a Wall-Modelled Large Eddy Simulation (WMLES)–based comparison of three intra-aortic pump designs in an aorta-sized pipe, and (ii) introducing a new dimensionless Hemolytic Number (HN) that enables standardized comparison of hemocompatibility across different pump configurations and operating conditions.

## 2. Methods

### 2.1. Intra-aortic pumps: design and structure

3D models of three different intra-aortic blood pumps were generated using computer-aided design (CAD) software, Solidworks (Dassault Systèmes, France). As the actual designs were not available to the authors due to proprietary restrictions, certain assumptions and design choices were made to approximate the pumps currently under development. It is therefore important to emphasize that the present study does not aim to reproduce the exact designs, but rather to explore pump configurations inspired by the Aortix™, ModulHeart™ and Second Heart Assist devices. The single-pump and triplet-pump geometries were created using literature-based reverse engineering in CAD, inspired by the HeartMate II device, as idealized axial-flow pumps. No CT scanning or direct physical measurement of a commercial pump was carried out. Instead, overall pump dimensions (total length, outer diameter), hub diameter, blade length, and blade angles were taken from previously published descriptions and engineering drawings of HeartMate-II–type axial pumps [31, 32]. These parameters were used to construct a HeartMate-II–like rotor–stator assembly in SolidWorks, preserving the main flow-driving features (rotor, hub, shroud, and blade orientation) while simplifying very small geometric details (e.g., local fillets, minor chamfers, internal filigrees) to ensure robust mesh generation. The triplet configuration was obtained by placing three identical copies of this single pump in series with a fixed axial spacing, without altering the internal blade channel geometry between modules by scaling down the dimensions of the single pump based on the main pump diameter, from 6 mm to 4 mm. Three 4 mm pumps were then positioned with equal radial angles (120°) between them. For the impeller-driven device, the blade span is 14.5 mm, and the outer diameter of the open stent is 22 mm. All three intra-aortic pump geometries were implanted inside a 22 mm simplified pipe, consistent with the diameter of the descending aorta. The impeller-driven pump was designed using a NACA 6509 airfoil for the blade geometry and was inspired by the Second Heart Assist circulatory assist device [28]. The detailed design features and dimensions of the geometries are displayed in **Figure 1**.

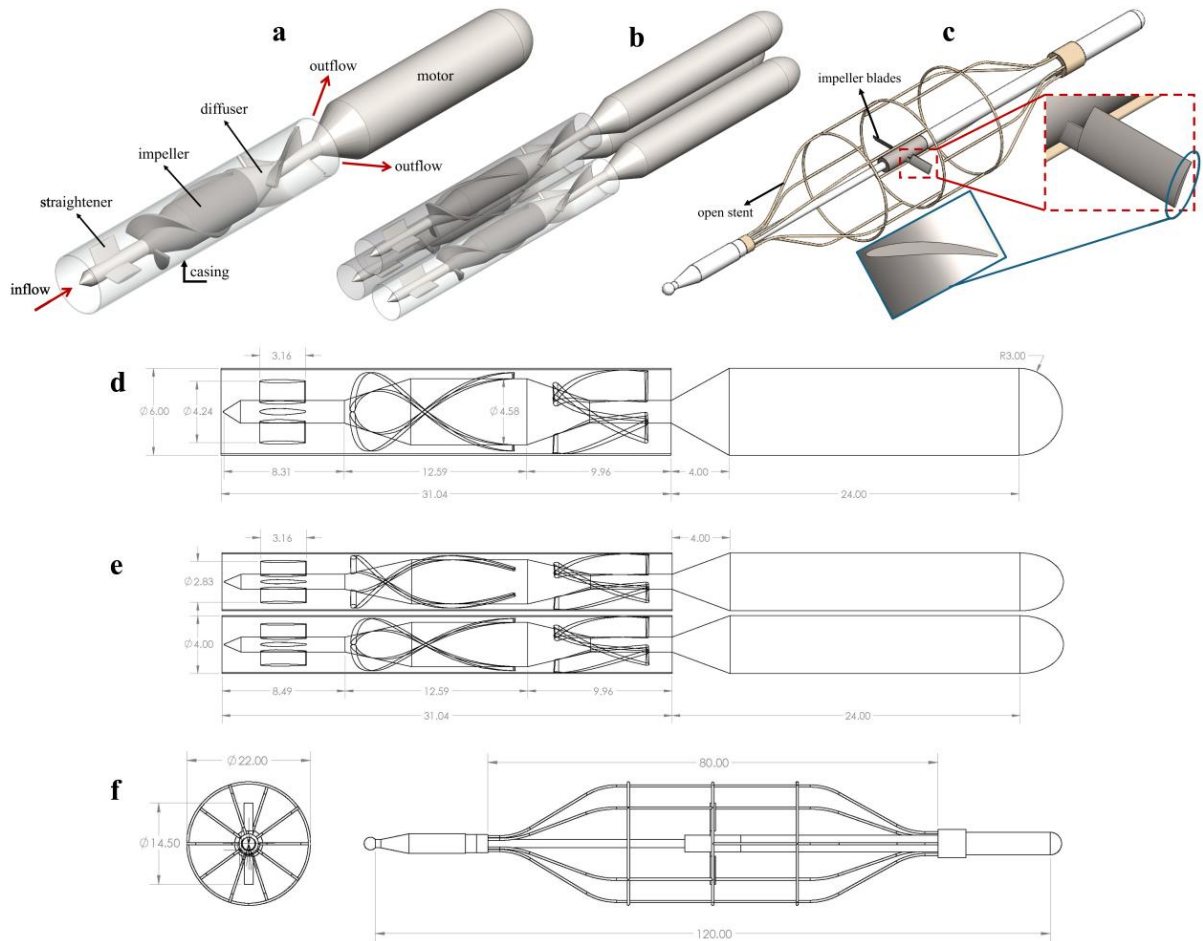

**Figure 1.** Design features and analysis configurations of the intra-aortic pumps. Isometric view; **(a)** single pump, **(b)** triplet-pump, **(c)** impeller-driven pump. The details of the dimensions are shown in **(d)** single pump, **(e)** triplet-pump, **(f)** impeller-driven pump. Dimensions are in millimeters.

***Design objectives and common constraints:*** The three intra-aortic pumps investigated in this study are intended as representative concepts for temporary intra-aortic flow augmentation. Because these devices operate non-occlusive within the aortic lumen, their design is governed by strong geometric and deployment constraints (limited outer diameter and device length) and by the need to balance flow augmentation against hemocompatibility, since shear exposure and wake/recirculation losses can increase blood damage risk. To enable a consistent, apples-to-apples comparison, all three concepts were constrained to the same intra-aortic diameter requirement and were evaluated within the same computational framework and operating-point definitions **(Table 1)**.

Table 1. Overview of simulation features and boundary conditions.

| Description | Conditions / Values |
|---|---|
| **Solver type** | Pressure-Based |
| **Time** | Transient |
| **Time-step** | $\Delta t = 0.033$ ms for 5° rotation (single pump) |
| | $\Delta t = 0.060$ ms for 5° rotation (triplet-pump) |
| | $\Delta t = 0.079$ ms for 5° rotation (impeller-driven pump) |
| **Rotation speed** | 25,000 rpm (single pump) |
| | 14,000 rpm (each one of triplet pump) |
| | 10,500 rpm (impeller-driven) |
| **Viscous model** | Wall-Modeled Large Eddy Simulation (WMLES) |
| **Algorithm for pressure-velocity coupling** | PISO |
| **Solution method for discretization** | Second Order / Bounded Second Order |
| **Inlet BCs** | Flow inlet: |
| (*for the second phase of the simulations*) | 0 to 8 L/min (single pump) |
| | 0 to 1.75 L/min (triplet-pump) |
| | 0 to 18 L/min (impeller-driven pump) |
| **Outlet BCs** | Pressure outlet: |
| (*for the second phase of the simulations*) | P = 100 mmHg |
| **Wall** | Rigid, No-slip condition |
| **Fluid model** | Blood as a continuous phase |
| **Density of fluid** | 1060 kg/m³ |
| **Viscosity of fluid** | Carreau viscosity model [18, 33] |
| | $\mu = \mu_\infty + (\mu_0 - \mu_\infty)[1 + (\lambda \dot{\gamma})^2]^{\frac{n-1}{2}}$ |
| | high-shear viscosity ($\mu_\infty = 0.00345\ Pa \cdot s$) |
| | low-shear viscosity ($\mu_0 = 0.056\ Pa \cdot s$) |
| | the time constant ($\lambda = 3.313\ s$) |
| | the power-law index ($n = 0.3568$) |

***Rationale for the three pump concepts:*** (i) The single axial pump is used as a baseline intra-aortic micro-axial/entrainment architecture, representing a compact rotor-driven concept commonly explored for catheter-based intra-aortic support. (ii) The triplet axial pump (three identical axial modules in series) is designed to investigate whether distributing the required momentum addition across multiple rotors can reduce localized shear intensity per module and modify downstream wake/recirculation behavior, while maintaining comparable overall flow augmentation. (iii) The impeller-driven pump represents an alternative intra-aortic architecture with a different blade-loading and flow-generation mechanism; it was designed using an airfoil-based blade profile (NACA 6509) to examine how this concept compares to axial architecture in terms of hemodynamic performance and hemolysis-related metrics under the same intra-aortic constraints.

***Operating conditions used in the present study:*** Rotational speeds for each design were selected within the range relevant to intra-aortic rotary assist concepts and are reported in **Table 1**. For each pump, simulations were performed over multiple operating points by prescribing a set of inlet mass-flow rates spanning the feasible range for that pump at the selected rotational speed, enabling consistent comparison of performance and hemolysis metrics across designs as described in **Section 2.2.2**.

## 2.2. Numerical method

### 2.2.1. Grid study

The solution grids for the pump geometries were generated using Fluent Meshing v22 (ANSYS Inc., Canonsburg, PA). The meshes consist of polyhedral elements and prism boundary layers along the wall, and local mesh refinement regions at the entrance of the pumps (for example, single and triplet-pump). The advantage of using polyhedral elements has been demonstrated in a previous study [34]. The present turbulence-resolving simulations were performed using WMLES, instead of wall-resolved LES. Accordingly, the near-wall mesh was designed to operate in the wall-model range rather than resolving the viscous sublayer, and the final meshes maintained $30 < y^+ < 60$ in the relevant pump passages and near the housing [35]. Polyhedral elements were employed to reduce numerical diffusion and achieve stable solutions with fewer control volumes compared with tetrahedral meshes.

For the mesh independence study, five distinct cases with different cell sizes were generated for the simulation of each pump geometry. Cell sizes for polyhedral elements were reduced with a refinement factor of 1.3 to increase the mesh density [36]. The mesh set used in the grid study spanned up to approximately 4–5 million cells per configuration, and the final production meshes (~1.1–1.7 million cells depending on the pump) were selected based on the Grid Convergence Index (GCI) criterion and computational feasibility. The computational domain was extended by 10 tube diameters (10D) upstream and 10D downstream of the pump to minimize inlet/outlet boundary influences on the developing rotor–wake flow and to prevent interaction of downstream recirculation structures with the outlet boundary condition. As shown in **Table 1**, the grid studies were performed under the condition of equal zero-total pressure at the inlet and equal zero-gauge static pressure at outlet ($P_{\text{inlet}} = 0\ Pa$, $P_{\text{outlet}} = 0\ Pa$). In this baseline configuration, no external pressure gradient is imposed across the domain; instead, flow is generated solely by the rotation of the pump, which adds momentum to the fluid and produces a local pressure rise across the pump. The resulting volumetric flow rate at the inlet represents the maximum flow capacity of each pump at the given rotational speed. To determine the optimal cell sizes that balance accuracy and computational cost, a grid study was performed using the GCI method [37, 38], which quantifies numerical uncertainties arising from spatial discretization and iterative convergence errors.

In this study, the variable of interest $f$ used in the Grid Convergence Index (GCI) analysis was the maximum volumetric flow rate generated by each pump configuration under these conditions. This quantity is central to the pump performance curves and is directly related to the pressure–flow characteristics used throughout this paper. Uncertainty and error bands were set to a maximum of 2%, ensuring that the pump's performance could be evaluated with high accuracy using WMLES. To achieve this, the mesh cases for each pump geometry were kept below the 2% error threshold. According to the results of the GCI analysis, the selected mesh cases have 1.21, 1.67, and 1.15 million elements for the single, triplet and impeller-driven pumps, respectively. All the details of the grid study using the GCI method and selected mesh cases for each pump are provided in the **Supplemental Material A (Tables A1–A4 and Figure A1)** section.

### 2.2.2. Numerical scheme and boundary conditions

The transient incompressible Navier–Stokes equations (momentum and continuity equations) were solved using a pressure-based finite-volume solver using ANSYS Fluent v22.1 (ANSYS Inc., Canonsburg, PA), with pressure–velocity coupling achieved via the Pressure-Implicit with Splitting of Operators (PISO) algorithm [39]. The PISO solver is specifically designed for transient incompressible flows and performs multiple pressure-correction steps within each time step, which strengthens the coupling between the pressure and velocity fields and improves stability for strongly unsteady flows. In the present pump–aorta configuration, the flow is three-dimensional, rotational, and highly unsteady, with strong rotor–stator interaction, recirculation, and wake dynamics. Under these conditions, PISO is widely used in LES and WMLES simulations because it offers robust convergence and computational efficiency for time-accurate incompressible flows in complex geometries [40–42]. To discretize the momentum equations, a second-order upwind scheme was utilized. The convergence criteria for both continuity and velocity were met when the residuals fell below $10^{-5}$.

As seen in **Figure 1a and 1b**, the straightener and diffuser were designated as stationary wall conditions for single and triplet pumps, while the impeller was defined as a moving wall boundary condition. The operating conditions were selected to reflect the intended working envelopes of representative commercially developed intra-aortic assist concepts (Aortix-, ModulHeart-, and SecondHeartAssist-like devices). Consequently, rotational speeds and the corresponding investigated inlet mass-flow ranges differ between designs. For each pump, a series of simulations was conducted across multiple inlet mass-flow operating points at the selected rotational speeds to generate pressure–flow characteristics and hemolysis metrics within the intended operating envelope. The pumps' rotational speeds are 25,000 rpm [2] and 14,000 rpm [21] for single and triplet pumps, respectively. In addition, time steps were chosen approximately as 0.033 ms and 0.06 ms to ensure 5° impeller rotation [32]. The rotational speed of the impeller-driven pump is 10,500 rpm, and the time step for the analysis of this pump was taken as 0.079 ms. The blade surfaces and the main rotating body (the dark-colored region seen in **Figure 1c**), which connects the blades, were defined as a rotating wall, and the rest of the pump's surfaces were defined as a stationary wall. The rotor motion was modeled using a transient sliding-mesh approach. The pump rotor subdomain was defined as a rigidly rotating zone and coupled to the stationary outer domain via a non-conformal rotor–stator interface (sliding interface). The time-step size was selected based on blade-passing resolution and CFL control in the pump passages. Specifically, because the present rotors have two main blades, 72-time steps per revolution corresponds to 36-time steps per blade-passing interval. The time step was further chosen to maintain a maximum CFL number below 0.85 across the investigated mesh levels. This approach is consistent with temporal-resolution practices reported for rotary blood pumps [19], where the time-step selection is guided by blade-passing resolution and CFL constraints. The simulations were carried out in two phases. In the first phase, to determine the maximal pump capacity for the grid independence analysis, both the inlet and the outlet pressures ($P_{inlet}$ and $P_{outlet}$) were set to 0 Pascal. The zero-pressure-gradient configuration is used exclusively to determine pump capacity and to perform the mesh-independence analysis, whereas the pressure–flow characteristics presented in the Results are obtained from simulations with prescribed mass flow rates and a pressure outlet boundary condition in the second phase. Detailed information about the simulation parameters, the defined boundary conditions and the non-Newtonian fluid used are presented in **Table 1**.

In the present configuration, the local Reynolds numbers inside and immediately downstream of the intra-aortic pumps are relatively high due to the small hydraulic diameters (4–6 mm) and the high rotational speeds. Using the pump hydraulic diameter and the mean velocity associated with flow rates up to 18 L/min, the characteristic Reynolds number in the pump passages is on the order of $Re \approx 4 \times 10^3 – 2 \times 10^4$. This indicates a transitional-to-turbulent regime rather than purely laminar flow and motivates the use of a turbulence-resolving approach in the pump and wake regions.

### 2.2.3. LES model

This study uses the Wall-Modelled Large Eddy Simulation (WMLES), which allows coarser meshes to be used in the near-wall region than in fully wall-resolved Large Eddy Simulation (LES). The complex geometry of the rotating pump, the narrow passages and the aorta-sized pipe would make wall-resolved LES prohibitively expensive at the considered Reynolds numbers, whereas RANS models are known to under-predict unsteady three-dimensional flow structures, such as rotor–stator interaction, tip-leakage vortices and recirculation zones, in rotary blood pumps. WMLES combines the benefits of LES and wall modelling techniques. It is a hybrid turbulence-modelling approach that provides a balance of accuracy and computational cost in evaluating critical parameters such as blood damage and pump performance. The large-scale eddies are resolved in the outer region of the boundary layer by the computational grid using a sub-grid-scale (SGS) model, while the eddies in the near-wall region are under-resolved and the wall shear stress is computed using a wall model [35, 43]. More details on governing equations of LES can be found in the **Supplemental Material B** section.

### 2.3. Evaluation of hemodynamic and hemolytic performance

### 2.3.1. Hemodynamic performance

In general, pressure head and flow rate are the main parameters to evaluate the performance of a pump, especially for analyzing intra-aortic blood pumps.

As previously explained, in the first stage of this study, the maximum capacity of each pump was investigated in terms of its ability to generate flow when the pumps were placed in a pipe, with the inlet and outlet pressure defined as 0 Pa. The maximum flow rates generated at specific rotational speeds for the single, triplet, and impeller-driven pumps were measured as 8.92 L/min, 2.09 L/min, and 20.09 L/min, respectively. Based on their maximum capacity, these pumps were subjected to different flow rates at the inlets, as shown in **Table 1**, to measure pressure head, hydraulic forces, and torque.

Hydraulic efficiency is another important parameter related to the pump's performance. It provides information about the cost-effective performance, such as energy consumption and heat production, of the pumps. The hydraulic efficiency for each pump was calculated as follows:

$$\eta = \frac{\dot{m}}{\rho}\left(\frac{P_{outlet} - P_{inlet}}{T\omega}\right) \tag{1}$$

where $\eta$ is hydraulic efficiency, $\dot{m}\ [kg/s]$ is mass flow rate, $P\ [Pa]$ is pressure, $T\ [Nm]$ is the mechanical torque acting on the impellers, and $\omega\ [s^{-1}]$ is the angular velocity, calculated from the rotational speed $n\ [rpm]$ using $\omega = \pi n/30.$,

### 2.3.2. Shear-induced hemolysis model

In this study, the hemolytic performance of the pumps was evaluated by calculating the hemolysis index (HI) and the normalized hemolysis index (NIH). For this, the scalar shear stress (SSS) in the three-dimensional fluid domain is calculated using Bludszuweit's stress formula [44] and the exposure time was evaluated using the Discrete Phase Model (1500 ± 200 massless particles were injected into the fluid domain every 10-time steps and tracked throughout the simulations). For each particle $k$, the exposure time $t_{\text{exp},k}$ was defined as the travel time between its injection at the inlet and its exit from the computational domain, $t_{\text{exp},k} = t_{\text{exit},k} - t_{\text{inj},k}$. The average exposure time reported in the Results (**in Figure 7b**) was then obtained by averaging over all particles that passed through the pump, $t_{exp} = (1/N)\sum_{k=1}^{N} t_{exp,k}$.

The hemolysis index is calculated using the power-law model proposed by Giersiepen et al. [45]:

$$HI(\%) = \frac{dHB}{HB} \times 100 = C t_{exp}^{\alpha} \tau^{\beta} \tag{2}$$

where $dHB$ is the amount of free hemoglobin, and $HB$ indicates the total hemoglobin concentration (150 g/L for adults). In addition, the values of the empirical constants representing human blood, $C = 3.458 \times 10^{-6}$, $\alpha = 0.2777$, and $\beta = 2.0639$, used in the equation above were taken from the study of Ding et al. [46]. Although hemolysis model coefficients reported in the literature are often derived using animal blood, this is common practice in blood pump research because in vitro and in vivo validation studies frequently rely on animal models and animal-derived blood. Prior work has also examined the sensitivity of power-law hemolysis models to blood type. In particular, Ding et al. [46] compared human, ovine, porcine, and bovine blood and reported that, for short exposure times and clinically relevant shear-stress levels, hemolysis indices are very similar across blood types. In the present simulations, the maximum exposure time is <0.05 s and the volume-averaged scalar shear stress remain <32 Pa across the investigated operating points; therefore, using the human-blood coefficient set reported by Ding et al. [46] is considered appropriate for the current comparative analysis. The normalized index of hemolysis (NIH), providing an estimate of the quantitative amount of hemolysis, is calculated using the following formula [12, 14, 47]:

$$NIH\ (g/100L) = 100 \times HI \times (1 - H_{ct}) \times HB \tag{3}$$

where $H_{ct}$ is the hematocrit level, and a value of 40% is selected in this study [30].

## 2.4. Validation Study

To strengthen confidence in the numerical framework, a validation study was performed by comparing CFD-predicted hydraulic performance curves against published experimental H–Q data for (i) an impeller-driven device (Whisper; Second Heart Assist) and (ii) an axial-flow rotary pump (Sputnik 2). These two datasets were selected because they provide openly reported pressure–flow measurements and represent the two flow-generation mechanisms examined in the present work (impeller-driven and axial-flow).

### 2.4.1. Reference datasets and operating conditions

*Whisper (impeller-driven):* Experimental H–Q curves for Whisper are reported at 8000 and 10,000 rpm using a glycerol–water blood-analog solution [48]. Because the impeller-driven pump investigated in the present study is operated at 10,500 rpm based on manufacturer specifications, while the literature dataset reports up to 10,000 rpm, we performed additional CFD simulations at 8000 rpm to enable a direct rotational speed-matched comparison and used the 10,500 rpm curve only for qualitative comparison against the nearest available experimental curve (10,000 rpm).

*Sputnik 2 (axial):* Experimental H–Q curves for Sputnik 2 are reported over multiple operating points, and the associated numerical setup in the reference study [49]. We selected the highest operating condition (9000 rpm) to perform validation analysis, because the operating condition of the single and triplet-pumps in this study are higher than reported experimental data. The single and triplet intra-aortic axial pump geometries investigated in this study were developed as HeartMate II–inspired representative axial-flow concepts. However, detailed HeartMate II CAD geometry and full bench-characterization datasets are not publicly available due to commercial restrictions, which limits direct device-specific validation. Therefore, to validate the axial-pump CFD methodology, we selected the Sputnik 2 axial-flow pump, for which experimental H–Q curves and key operating conditions are available in the literature. This provides a practical validation of the axial-flow pump modeling workflow used in the present study.

### 2.4.2. Validation results

**Figure 2** compares the CFD-predicted H–Q curves to the corresponding experimental curves. For the Whisper dataset, the 8000 rpm CFD curve reproduces the experimental trend and magnitude over the investigated flow range, capturing the expected decrease in pressure rise with increasing flow rate. For 10,500 rpm, the CFD curve is compared qualitatively to the nearest available experimental curve at 10,000 rpm; the predicted pressure-rise level remains in the same range, while differences in slope are expected given the speed mismatch and unreported test-loop details. For Sputnik 2 at 9000 rpm, the CFD predictions reproduce the characteristic monotonic reduction in pressure rise with increasing flow rate and show close agreement in both slope and magnitude across the experimental operating points. Overall, these comparisons support that the present numerical framework provides physically consistent hydraulic performance trends for both impeller-driven and axial-flow rotary pump configurations.

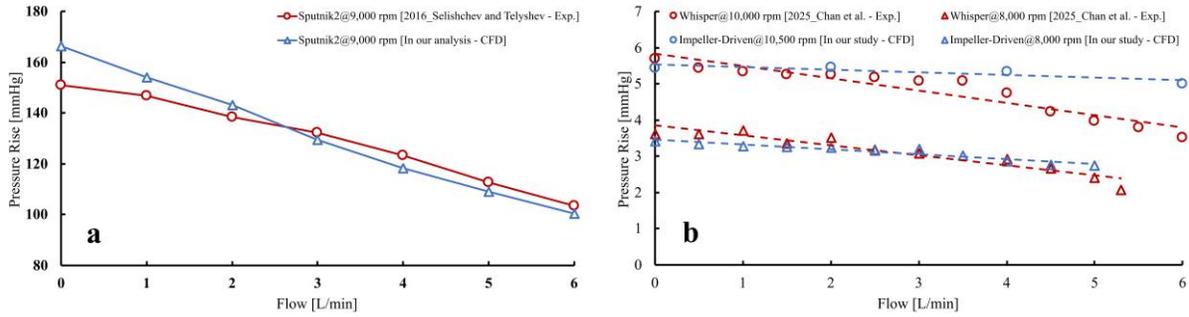

**Figure 2.** Validation of CFD-predicted hydraulic performance against published experimental H–Q curves. (**a**) Whisper (Second Heart Assist) comparison: experimental curves at 8000 and 10,000 rpm from Chan et al. (2025) and CFD predictions for the impeller-driven pump at 8000 rpm (speed-matched) and 10,500 rpm (nearest-speed qualitative comparison) (2025 - Chan et al). (**b**) Sputnik 2 comparison: experimental H–Q curve at 9000 rpm from Selishchev and Telyshev (2016) and CFD prediction for the Sputnik 2 geometry at 9000 rpm using the same flow-sweep approach.

## 3. Results

A baseline study was first performed to provide insight into the subsequent hemodynamic analysis of three intra-aortic blood pump configurations: single, triplet and impeller-driven. As shown in **Figure 1**, the design characteristics of the single and triplet pumps differ from those of the impeller-driven pumps. Although they share similar design features, the single and triplet pumps differ in the number of impellers, as well as pump size and rotational speed. The flow capacity of the pumps depends heavily on their rotational speed ($\omega$) and diameter ($D$), as described by the equation $\dot{Q} = f(\omega, D)$. Therefore, the flow capacity differs between the single and triplet pumps.

The results are displayed in **Figure 3** and show that the flow rate increases quite linearly with the rotational speed and non-linearly with the diameter. Considering standard values used in the literature for intra-aortic pumps currently under development, the flow rate generated for a 6 mm-diameter pump operating at 25000 rpm can generate a maximum of 8.59 L/min. For the triplet pump configuration, three 4 mm-diameter pumps rotating each at 14000 rpm can generate a maximum of 2.13 L/min (0.71 L/min for each pump). Again, note that in this configuration the flow is generated solely by the rotation of the pump, with no pressure gradient imposed across the pipe.

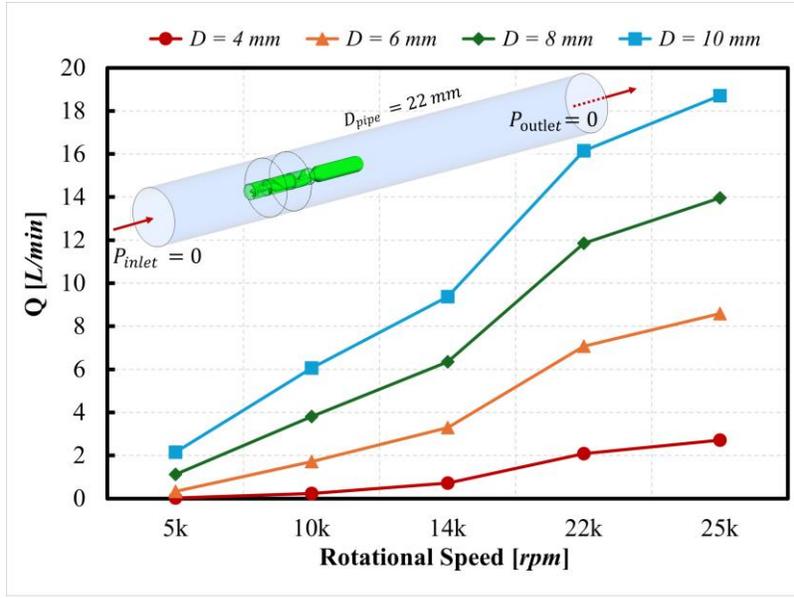

**Figure 3.** Variation in the flow rate generated by different pump diameters and rotational speeds. Note that here the inlet and outlet were set to 0 Pa, and the volumetric flow is generated solely by the rotational speed of the pump.

### 3.1. Characteristic curves and hydraulic efficiency

**Figure 4** shows the hydraulic performance curves of the three different intra-aortic pump configurations considered in this study. The figure presents the correlation between volumetric flow rate, pressure rise, and hydraulic efficiency on the left-hand side, and the behavior of torque and axial force on the right-hand side.

For the single pump, the measured pressure rise shows a distinct decrease as the flow rate increases, dropping from about 170 Pa at 0 L/min to almost 25 Pa at 8 L/min. At the same time, hydraulic efficiency follows a downward curve, decreasing from approximately 2.7% to less than 2.0%. The torque initially displays a small increase before a subsequent decrease, while the axial force steadily increases as the flow rate rises. It is also observed that, in the red-shaded region of 0–2 L/min, backflow (i.e. flow recirculation between the inlet and the outlet of the pump) occurs. Under such conditions, the pump has to recirculate the flow from its outlet back to its inlet to compensate for the imbalance between the inflow and the pump's set flow rate.

The triplet pump shows a similar, though less pronounced, trend. The pressure rise decreases from about 50 Pa to 30 Pa as the flow rate increases to 1.75 L/min. Hydraulic efficiency changes from 2.17% down to 1.38%, reflecting moderate sensitivity to changes in flow rate. The torque exhibits a weak nonlinear relationship with the flow rate, while the axial force decreases approximately linearly as the flow rate increases. Similar to the single pump, the red-shaded area (up to about 0.75 L/min) represents the flow conditions for which the triplet-pump experiences a recirculation of the flow from its outlet to its inlet.

The performance of the impeller pump shows very distinctive features. It reaches a very high-pressure rise of about 800 Pa when the flow rate is around 4 L/min, before decreasing to about 400 Pa at 18 L/min. Hydraulic efficiency roughly follows the flow rate increase, reaching about 6.0% at a flow level of approximately 14 L/min before slightly decreasing. Both torque and axial force follow parabolic trends, with both quantities peaking at intermediate flow levels, 8 to 10 L/min before declining, showcasing potential optimal operating conditions in terms of torque, axial force and generated flow. No recirculation of the flow can be observed with the impeller-driven pump.

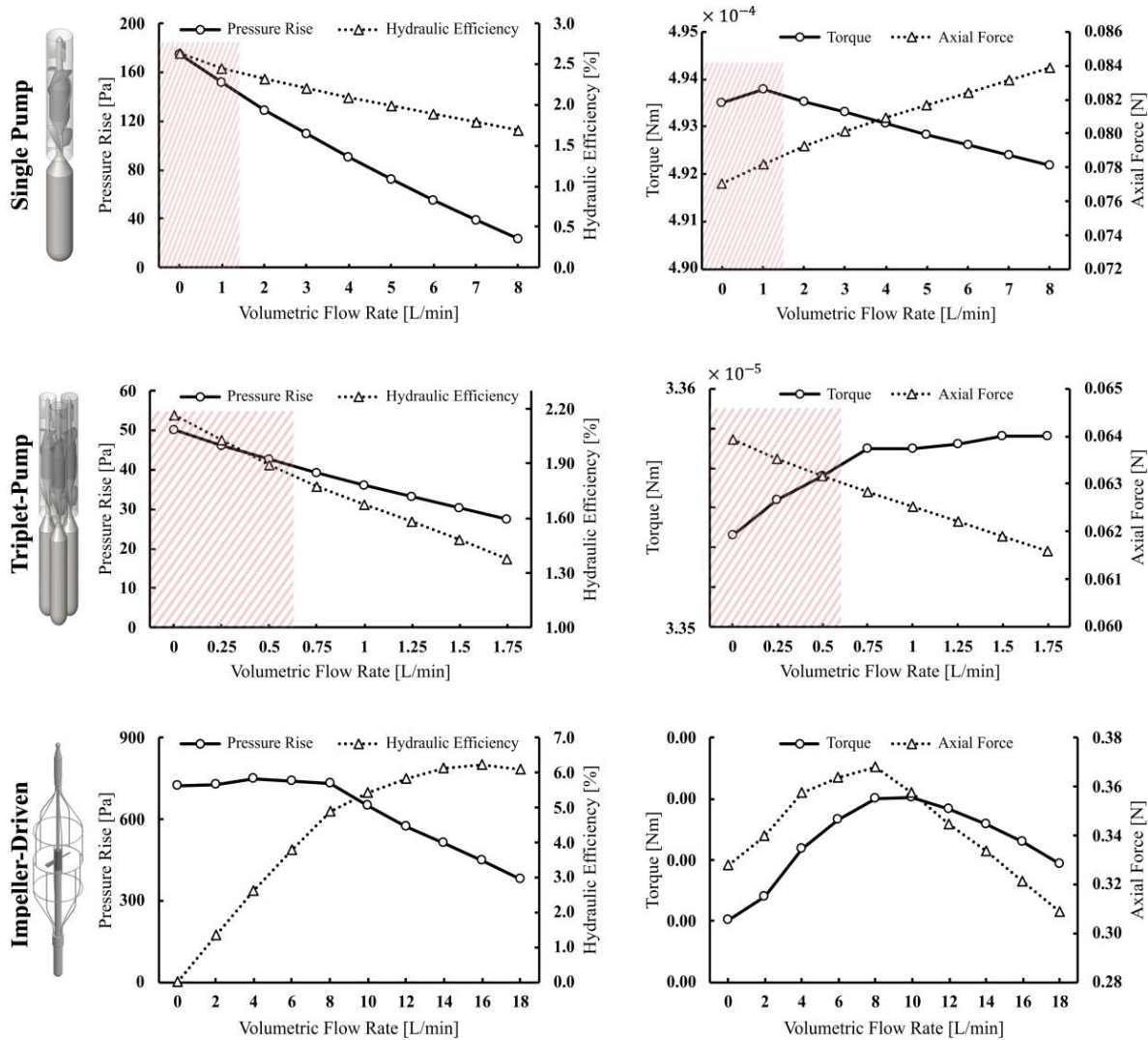

**Figure 4.** Hydraulic performance curves of the intra-aortic pumps (single pump, triplet-pump and impeller-driven pump). *Left:* pressure rise and hydraulic efficiency, *Right:* torque and axial force. The *Red-Shaded Regions* indicate the range of volumetric flow rates for which recirculation occurs from the pump outlet to the inlet for the single and triplet pumps (see text for details).

## 3.2. Flow fields generated by intra-aortic pumps

**Figure 5** shows the velocity distribution for each pump configuration under the described flow and rotational speed conditions. As seen in **Figures 5a–c**, the triplet pump in **Figure 5b**, running at 14,000 rpm and a total flow rate of 1 L/min, has a more regular and well-organized velocity profile compared to the single pump in **Figure 5a** and the impeller-driven pump in **Figure 5c**. The velocity patterns of the single and impeller-driven pumps exhibit more complex and turbulent flow structure downstream of the pumps. The single pump experiences the highest velocities, reaching up to 3 m/s, whereas the triplet pump reaches less than 1.5 m/s.

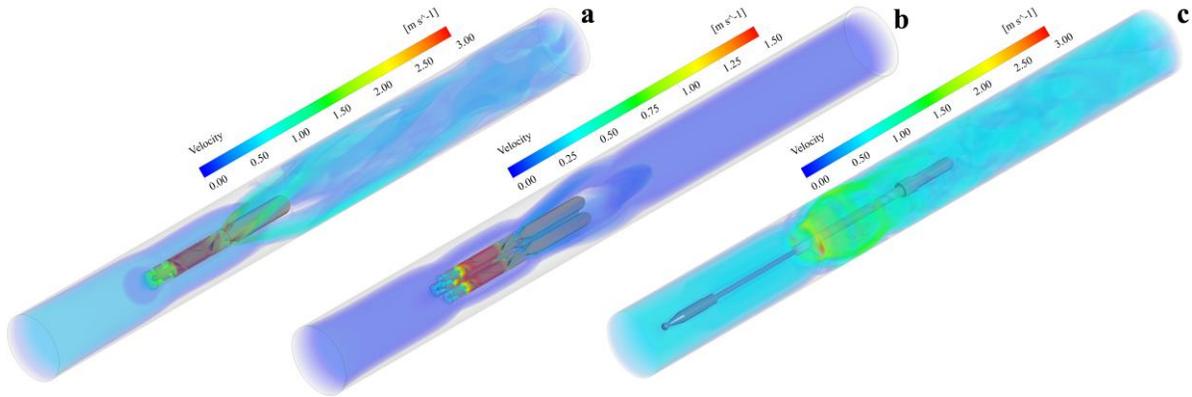

**Figure 5.** Three-dimensional velocity distribution. (**a**) Single pump at 25000 rpm and 5 L/min, (**b**) triplet pump at 3x14000 rpm and 1 L/min, (**c**) impeller-driven pump at 10500 rpm at 10 L/min

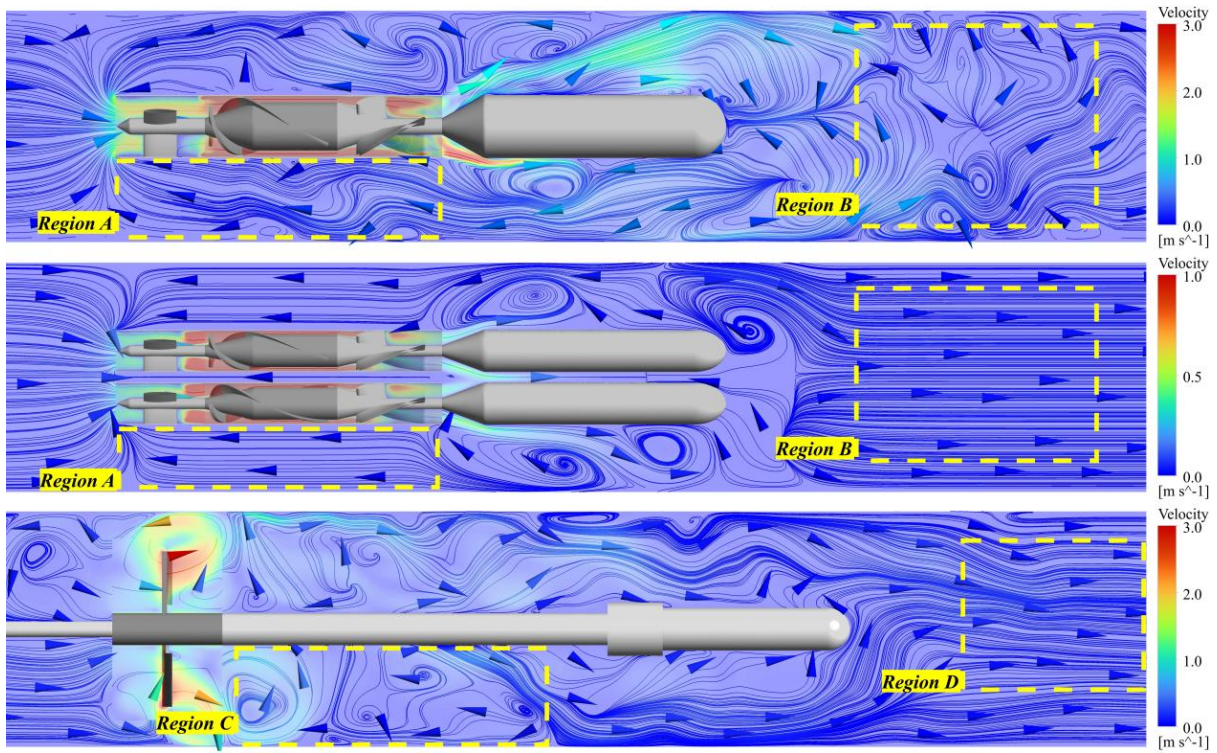

**Figure 6.** Velocity Streamlines for low flow conditions, corresponding to retrograde flow for the single and triplet-pump. *Top*: single pump at 1 L/min, *Middle*: triplet-pump at 0.25 L/min, *Bottom*: impeller-driven pump at 2 L/min.

**Figure 6** specifically highlights the recirculation flow patterns induced by the single and triplet-pump at low flow conditions (corresponding to the red-shaded regions in **Figure 4**). Indeed, a strong retrograde flow can be observed around the pump casings, from the outlet to the inlet of the pumps as shown in *Region A*. This phenomenon, caused by a lower flow rate at the pipe inlet than the pump capacity, increases the risk of long residence time and thrombogenicity, as described in Georges' study on the ModulHeart intra-aortic pump [50]. Because of the difference in the rotational speed and design between the different pumps, distinct flow patterns can be observed downstream of each configuration. The velocity profiles appear to be more uniform for the triplet pump design when compared to the single pump (*Region B)*. In the case of an impeller-driven pump, two distinct regions can be observed: 1) a near wake region (Region C) just downstream of the impeller with complex flow structures and 2) a far wake region (region D) where the flow is getting more uniform.

### 3.3. Blood damage analysis

**Figure 7** shows comparative profiles related to the severity of blood damage caused by the single, triplet, and impeller-driven pump configurations, plotted as scalar shear stress (SSS), volume average exposure times, and hemolysis index (HI). Each pump configuration was simulated at multiple prescribed inlet mass-flow rates spanning its operating range. **Figure 7** summarizes the resulting NIH and HN values across these operating points using a box-and-whisker representation; therefore, the whiskers/boxes indicate the spread across simulated operating conditions. As depicted in **Figure 7a**, the single pump has the highest percentage of fluid volume (49.14%) under SSS levels greater than 9 Pa, followed by the triplet-pump at 31.47% and the impeller pump at 26.69%. Regarding the >50 Pa level, the single pump has the highest ratio with 31.47%, while the triplet and impeller-driven pumps show a very low ratio with 3.07% and 2.67%, respectively. The same pattern is observed at SSS level of >150 Pa—which are strongly associated with higher levels of hemolysis—where the single pump reflects the highest percentage at 12.1%, significantly ahead of the triplet pump at 0.15% and the impeller-driven pump at 0.21%. In terms of mean exposure time (see **Figure 7b**), the single pump and the triplet pump have relatively similar exposure times of 0.0303 s and 0.0396 s, respectively: in comparison, the impeller-driven pump has a significantly shorter exposure time of 0.0072 s. The reduction in exposure time increases hemocompatibility by reducing the duration during which blood cells are subjected to shear forces. The NIH values in **Figure 7c** reinforce these findings even further. The data for the single pump reveal an HI of about 0.00159% and an NIH of about 0.1434 g/100L. By comparison, performance is improved in the case of the triplet pump, with HI dropping to about 0.00018% and NIH to 0.0162 g/100L. Also, a steady reduction in HI and NIH is observed as the flow rate increases across the impeller-based pump. Starting from around 0.000245% HI, this steadily decreases as the flow rate increases, reaching a minimum of 0.000039% at 18 L/min. The NIH also decreases significantly from over 0.0155 g/100 L to 0.0035 g/100 L.

One limitation of the NIH definition is that it is not a dimensionless and does not provide insight into the balance between the different effects in the complex flow configuration induced by intra-aortic pumps. Therefore, a new dimensionless parameter was introduced to compare the hemolytic performance of different pumps under varying operating conditions. This parameter has been derived using the Buckingham Pi theorem. The proposed dimensionless number is called *Hemolytic Number (HN)* and the resulting expression can be written as

$$\prod_{HN} = C_E \cdot \frac{\tau \cdot t_{exp}}{\rho \cdot \omega \cdot D^2} \qquad (4)$$

where $C_E = 1000$ is a scaling constant, $\tau$ is the scalar shear stress, $t_{exp}$ is the exposure time, $\rho$ is the fluid density, $\omega$ is the angular velocity, $D$ is the pump diameter. Physically, HN reflects the ratio of the blood damage index to the characteristic inertial stress. By integrating this dimensionless number, the study introduces a standardized metric for comparing hemolysis of the different blood pumps.

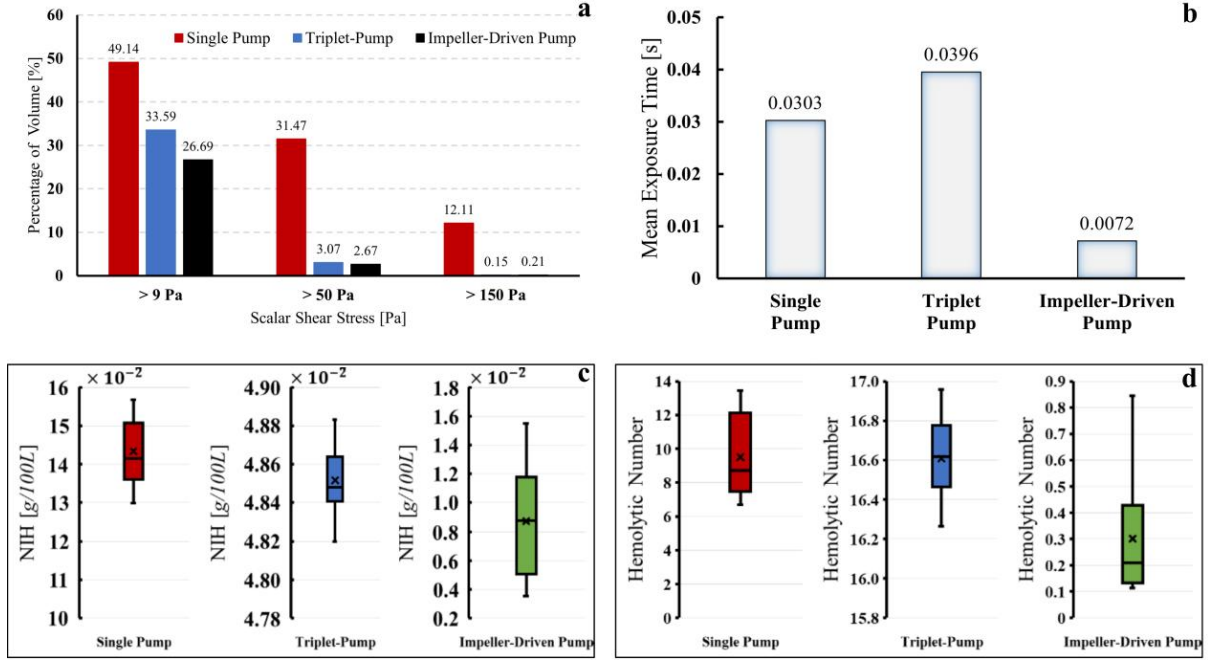

**Figure 7.** Blood damage–related metrics for the three pump configurations. (**a**) Percentage of fluid volume exposed to different thresholds of scalar shear stress (SSS), (**b**) Mean exposure time for each pump configuration, (**c**) Distribution of normalized index of hemolysis (NIH) across the simulated operating points (different prescribed inlet mass-flow rates), (**d**) Distribution of the Hemolytic Number (HN) across the same operating points. For panels (**c**) and (**d**), the whiskers indicate the minimum and maximum values obtained among the simulated cases; the colored box spans from the second smallest to the second largest value (i.e., excluding the extreme min/max); the black horizontal line inside the box denotes the median; and the "×" marker denotes the mean. These are compact representations of the variation of hemolysis metrics with operating conditions.

As shown in **Figure 3**, the flow rate generated by the intra-aortic blood pumps depends on the rotational speed and pump diameter. The amount of blood damage is closely related to these parameters, as well as to the design features and parameters. The desired performance of blood pumps is to achieve a high flow rate and low hemolysis. Accordingly, the smallest $NIH/Q$ ratio should be determined to identify the pump with the best hemocompatibility performance. However, this ratio provides meaningless units for a reasonable comparison between pumps, and the literature lacks parameters for evaluation, to the authors' knowledge. In accordance with this requirement, a dimensionless number has been proposed based on the Buckingham Pi theorem. The detailed Buckingham–Pi dimensional analysis leading to this expression is provided in **Table C1** from **Supplementary Material C**. Moreover, **Figure 8** shows the relationship between $NIH/Q$ and $HN$ to evaluate the accuracy of this trend.

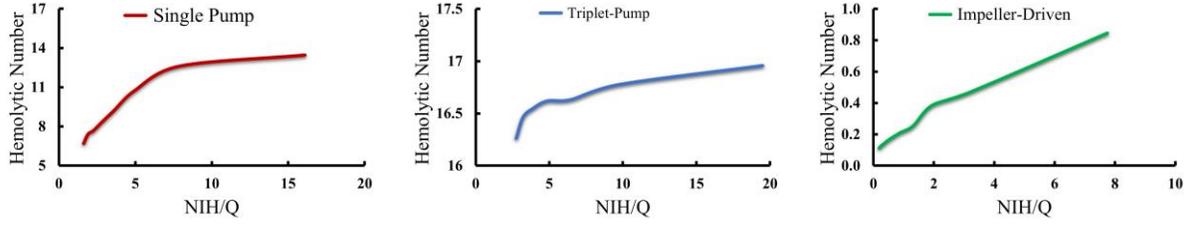

**Figure 8.** The evaluation of the Hemolytic Number for each pump at different flow rates.

**Figure 7d** shows the HN values calculated for each pump at different flow rates. The smallest HN value represents the pump that is most effective in terms of the relationship between flow rate and blood damage. According to these results, the impeller-driven pump has the best hemocompatibility performance, with HN values consistently below 1 over the investigated flow range, in contrast to the single and triplet pumps.

The regions of the flow domain associated with elevated risks of hemolysis, as determined using the HI, for each pump design are depicted in **Figure 9.** The HI distributions presented in **Figures 9a-c** strongly reflect the observed flow characteristics. In the single pump case (**Figure 9a**), regions of high hemolysis levels are seen adjacent to the tips of the impeller blades, where HI levels exceed 0.20%, indicative of high local shear stress. The triplet pump (**Figure 9b**) distributes the resulting damage over a larger area, with a maximum HI of around 0.05%, thereby reducing peak levels of damage. The impeller pump (**Figure 9c**) shows the best hemocompatibility profile, keeping HI values under 0.05% across the entire flow area while also maintaining very small recirculation regions.

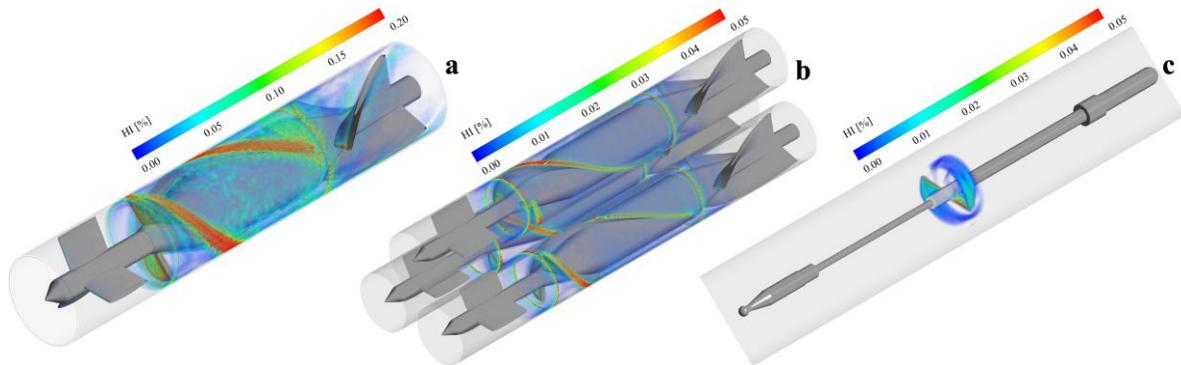

**Figure 9.** Volume rendering of hemolysis index distribution at different operating conditions. (**a**) Single pump at 25000 rpm and 5 L/min, (**b**) triplet pump at 3x14000 rpm and 1 L/min, (**c**) impeller-driven pump at 10500 rpm and 10 L/min

## 4. Discussion

This study represents a comprehensive computational comparison of three intra-aortic blood pump configurations that are currently under development: a single pump, a triplet-pump, and an impeller-driven pump. The emphasis was placed on evaluating their hemodynamic performance and hemolytic potential. The key findings provide meaningful insights into the design and optimization of intra-aortic circulatory assist devices, particularly with regard to balancing flow generation and blood compatibility: (1) intra-aortic blood pumps generally have low hydraulic efficiency; (2) for the single and triplet pumps, especially at low flow rates, recirculation can occur from the outlet to the inlet of the pump; and (3) pump design has a significant impact on the hemocompatibility performance of the device.

From a numerical-modelling standpoint, the use of WMLES in this work is motivated by the transitional-to-turbulent flow regime inside and downstream of the pumps and the need to resolve large-scale unsteadiness associated with recirculation and wake structures. While WMLES was originally developed for high-Reynolds-number external flows, it has increasingly been applied to internal and non-Newtonian flows [51, 52], where it effectively acts as a hybrid RANS–LES method: sub-grid activity remains low in laminar regions and increases only where instabilities develop. In the present configuration, this enables accurate prediction of the core flow structures governing shear stress and residence time, at a computational cost that would not be feasible for fully wall-resolved LES of the entire pump–aorta system.

A preliminary analysis of a single pump was performed by considering different sizes and rotational speeds. These simulation results highlight the synergistic role of geometric size and rotational speed in determining the pumps' hydraulic output. The flow generated increases almost linearly with rotational speed, but non-linearly with pump diameter. Such behavior is consistent with the work of Throckmorton et al [53] on axial flow pumps.

Due to their different design configurations (i.e., pump size, number of pumps) and operating conditions (particularly rotational speed), the three configurations have different flow-rate capacities. Therefore, they were analyzed over a wide range of flow rates, extending beyond strictly physiological values, in order to obtain their performance curves. In this study, we evaluated the hydraulic efficiency of these pumps and found that the impeller-driven pump exhibits higher efficiency (~6% at 14 L/min) compared with the single pump (2.7%) and the triplet-pump (2.2%). Clearly distinguish between LVAD hydraulic efficiencies (≈10–30%) and the lower efficiencies expected for intra-aortic micro axial pumps due to their different configuration and design constraints [12, 15]. Explicitly state that, to our knowledge, no prior studies have reported hydraulic efficiency for intra-aortic pumps, and that our values represent an initial estimate within a unified CFD framework. The global efficiency definition used here is intentionally conservative and naturally leads to lower numerical values than LVAD efficiencies but is appropriate for assessing the overall "assist effectiveness" of intra-aortic devices. Because the intra-aortic pumps considered here do not occlude the aortic lumen and operate by entraining native aortic flow, only a portion of the total aortic flow is accelerated and pressurized by the pump, while the remainder bypasses the device. When hydraulic efficiency is defined with respect to the pump cross-section, this non-occlusive, entrainment-based configuration naturally yields lower numerical values. In addition, the strong recirculation and wake regions observed around the pump body and in the downstream aorta introduce extra viscous dissipation, which further reduces the fraction of mechanical power converted into useful hydraulic work.

Regardless of the flow rate defined at the pipe inlet, the flow rate entering the pump casing remains approximately constant, because the pump's suction capacity is directly related to its rotation speed. As the inlet flow rate increases, the peripheral flow bypassing the pump also increases, creating additional resistance at the pump outlet. At low flow rates, when the inlet flow rate $\dot{Q}_{inlet@pipe}$ is lower than the pump capacity $\dot{Q}_{pump}$, recirculation occurs around the pump body, with flow returning from the outlet towards the inlet. This recirculation further reduces hydraulic efficiency, consistent with previous observations [50].

The hemolysis index decreases as the flow rate increases, while the impeller-driven pump consistently shows better hemocompatibility than the other two designs. This improvement can be attributed to its simple and compact configuration. It exhibits the lowest SSS distribution, which remains below hemolytic thresholds, an average exposure time of only 0.0072 s, and the smallest NIH (0.0035 g/100L) under the highest operating speeds.

Although the impeller-driven pump generates elevated shear zones near the blade tips, its overall hemolysis risk remains lower than that of the single pump because it minimizes large recirculation regions and long residence times. This is consistent with previous findings that retrograde flow and increased residence time play a major role in thrombogenicity and blood damage. The triplet pump shows intermediate performance: it significantly reduces maximum shear stress compared with the single pump but exhibits longer exposure times, resulting in lower hemocompatibility than the impeller-driven pump.

While the Normalized Index of Hemolysis (NIH) is widely used to quantify blood damage, it is not dimensionless and remains strongly dependent on device size and operating conditions, which complicates comparison across different intra-aortic pump designs. The proposed Hemolytic Number (HN) overcomes this limitation by combining shear stress, exposure time, density, rotational speed, and pump diameter into a single dimensionless parameter. *HN* thus provides a standardized measure of the hemolytic burden relative to inertial loading and flow generation. In this framework, lower HN values indicate more favorable hemocompatibility for a given level of hydraulic performance, enabling direct comparison between pump configurations and operating points.

There are several limitations to this study. All pump geometries used in this study are idealized, HeartMate-II–inspired models reconstructed from published global dimensions and design parameters rather than exact replicas obtained from CT or direct measurement of a commercial device. The simulations were performed using a rigid pipe representing the descending aorta, primarily to reduce computational costs and to isolate the effects of pump design on the performance curves of different intra-aortic blood pumps, given the geometric and hemodynamic complexities of the human aorta. In addition, a steady inlet boundary condition was prescribed, which facilitated precise evaluation of hemodynamic parameters but does not fully reflect physiological pulsatile flow. Future studies should extend this work by incorporating more physiological models, including patient-specific anatomies and pulsatile inlet boundary conditions. The proposed *Hemolytic Number* could also be evaluated further to establish meaningful ranges or threshold that may guide clinical decision-making alongside conventional circulatory assist devices. This can be pursued both numerically and experimentally by reproducing clinically relevant flow regimes.

Finally, WMLES remains an approximate turbulence-modelling approach, particularly in regions where the flow is weakly turbulent or transitional. Although the grid-convergence analysis and the qualitative agreement of flow features with previous studies support the present modelling choice, future work could explore alternative hybrid RANS–LES formulations or wall-resolved LES in localized regions to further validate the predictions.

## 5. Conclusion

This study shows that, despite the promising role intra-aortic blood pumps can play in supporting the left ventricle and augmenting renal flow, current intra-aortic pump designs exhibit relatively low hydraulic efficiency compared with conventional axial and centrifugal blood pumps. In particular, for coaxial configurations—such as single and triplet pumps—flow recirculation around the pump body can occur under low-flow conditions, further reducing efficiency and potentially increasing blood damage. The compact impeller-driven pump design demonstrates superior hemodynamic and hemocompatibility performance, combining higher pressure generation and flow capacity with lower scalar shear stress, shorter exposure times, and reduced hemolysis indices compared with the other designs. Finally, the newly proposed dimensionless Hemolytic Number (HN) provides a unified framework to compare different intra-aortic pump designs operating under varying conditions and may support future device optimization and clinical decision-making.

## Conflict of interest statement

No potential conflict of interest was reported by the authors.